\documentclass{cimento} 

\usepackage{ragged2e}
\usepackage{etoolbox}
\apptocmd{\frame}{}{\justifying}{}

\usepackage{xcolor}

\usepackage{amsfonts,amsmath,amssymb,bm,latexsym,mathrsfs}
\usepackage{booktabs,multirow}
\usepackage{graphicx}
\usepackage{epstopdf}
\usepackage{setspace}
\usepackage{subfigure}
\usepackage{pgfplots}
\usepackage[applemac]{inputenc}
\everymath{\displaystyle}
\usepackage[none]{hyphenat}

\title{Exotic spectroscopy in a diquark model with a little help from AdS/QCD} 
\author{F.~Giannuzzi\from{ins:x}}
\instlist{\inst{ins:x} INFN, Sezione di Bari - Bari, Italy}

\begin{document} 
\maketitle

\begin{abstract}
 Masses of heavy mesons, tetraquarks and pentaquarks computed in a potential model are shown. Tetraquarks are studied as bound states of a diquark and an antidiquark. Pentaquarks are constructed from a series of two-body interactions between the four quarks and the antiquark.
 New results on $Qs\bar Q \bar q$ and $QQ\bar Q \bar Q$ tetraquarks are also obtained.
\end{abstract}

\section{Introduction}
High-energy experiments are observing a lot of heavy resonances. Exotic tetraquark and pentaquark states have been discovered, and even states with quantum numbers of ordinary mesons and baryons but puzzling features could have an unconventional quark content. A feature one can look at in order to understand the internal structure of the observed states is the mass.

In the following, the masses of heavy mesons, tetraquarks and pentaquarks computed in a potential model \cite{Carlucci:2007um,Giannuzzi:2019esi} are shown and compared to experimental data. 
New results on $Qs\bar Q \bar q$ and $QQ\bar Q \bar Q$ tetraquarks are also shown, for which some candidates have been recently observed \cite{{pdg,LHCb:2021uow,LHCb:2020bwg}}.
The model is characterised by two main features: relativistic kinematics and the use of a quark-antiquark static potential developed in an AdS/QCD model \cite{Andreev:2006ct}. 
We have used the Salpeter equation describing a two-body interaction for finding meson masses and also for tetraquarks and pentaquarks, interpreting these states as the result of a series of two-body interactions: a tetraquark is studied as the bound state of a diquark and an antidiquark, a pentaquark as the bound state of an antiquark and a two-diquark state or assuming different combinations. Basically, in this way the problem of studying the interactions among three or more quarks has been reduced to solving a series of two-body interactions in a cascade \cite{Giannuzzi:2009gh}. 
Other models based on the interaction between quarks and diquarks or diquarks and antidiquarks are described e.g. in \cite{Maiani:2004vq,other,ferrettisantopinto,bedolla2020}.

\section{Model}
The Saltpeter equation is a wave equation describing a two-body interaction with relativistic kinematics. It reads:
\begin{equation}\label{eq:salpeter}
  \left(\sqrt{m_1^2-\nabla^2}+\sqrt{m_{2}^2-\nabla^2}+{V}(r)\right)
\psi({\bf r})\,=\,M\, \psi({\bf r})\,.
\end{equation}
$m_1$ and $m_2$ are the masses of the interacting particles, while $M$ and $\psi$ are the mass and wavefunction of the final bound state.
We consider only hadrons with at least one heavy quark, for which the assumptions involved in a potential model are better realised, and with zero orbital momentum ($\ell=0$).

The main term of Eq.~\eqref{eq:salpeter} is the potential $V(r)$. It contains three contributions: 
\begin{equation}
  V(r)=V_{QCD}(r) + V_{spin}(r) + V_0\,.
\end{equation}
To describe the strong interaction, we use the potential $V_{QCD}(r)$ computed through the AdS/QCD correspondence in \cite{Andreev:2006ct}, a bottom-up approach inspired by the gauge/gravity dualities. 
The potential is expressed in parametric form:
\begin{eqnarray}
  r(z_0)\,&=&\,2\, z_0  \int_0^1 dv\, v^{2} e^{c z_0^2 (1-v^2)/2} \left(1-v^4 e^{c z_0^2 (1-v^2)}\right)^{-1/2}\label{r}\\
V_{QCD}(z_0)\,&=&\,\frac{g}{\pi} \frac{1}{z_0} \left( -1+\int_0^1 dv\, v^{-2} \left[ e^{c z_0^2 v^2/2} \left(1-v^4 e^{c z_0^2 (1-v^2)}\right)^{-1/2}-1\right]\right)\,.
\end{eqnarray}

$V_{spin}(r)$ describes the spin-spin interaction. In the one-gluon-exchange approximation it can be written as  \cite{Barnes:2005pb}:
\begin{equation}
  V_{spin}(r)\,=\,A \frac{\tilde\delta(r)}{m_1 m_2}{\bf S_1}\cdot{\bf
S_2} \qquad\quad\mbox{with }\qquad \tilde\delta(r)=\left(\frac{\sigma}{\sqrt{\pi}}\right)^3
e^{-\sigma^2 r^2}\,,
\end{equation}
where $\bf S_i$ are the spin of the interacting particles. $A$ is a parameter getting different values for hadrons containing charm and bottom quarks ($A_c$, $A_b$) to account for the different values of $\alpha_s$ at the two scales. 
Finally, $V_0$ is a constant term (parameter).
We also add a cutoff at short distances to cure the singularity of the wavefunction, an artifact of the approximations involved to get the Salpeter equation: $V(r) = {V}(r_M)$ for $r\leqslant r_M$, with $r_M=k/M$ \cite{Colangelo:1990rv}. 

\section{Results}
The parameters of the model are fixed from the masses of mesons containing at least one heavy quark, charm or bottom.
Meson masses are obtained by solving Eq.~\eqref{eq:salpeter}, with $m_1$ and $m_2$ the masses of the constituent quark and antiquark.
The results are in Table~\ref{tab:mesons}, and the fitted parameters are \cite{Carlucci:2007um}: $c=0.300$ GeV$^2$, $g=2.750$, $V_0=-0.488$ GeV, $k=1.48$, $A_c=7.920$,  $A_b=3.087$, $\sigma=1.209$  GeV (in the potential), $m_q=0.302$ GeV,  $m_s=0.454$ GeV, $m_c=1.733$ GeV, $m_b=5.139$ GeV (constituent quark masses).
The input masses \cite{pdg} of the fit are denoted by a star in the table, and they are reproduced by the model with a mean error $\sim 0.63\%$.
The masses of $2S$ $c\bar q$ and $c\bar s$ states are compatible with those of the observed states $D_{0}(2550)$ and $D_{1}^*(2600)$, and $D_{s0}(2590)$ and $D_{s1}^*(2700)$. $D_1^*(2760)$ and $D_{s1}^*(2860)$ have instead too low masses to be the $3S$ $1^-$ $c\bar q$ and $c\bar s$ states, respectively. The masses of excited $c\bar c$ $1^{--}$ states can be compared with the experimental candidates observed so far, among which $\psi(3770)$, $\psi(4040)$, $\psi(4160)$, $\psi(4230)$, $\psi(4360)$, $\psi(4415)$, $\psi(4660)$.
In the bottom sector we support the hypothesis that $\Upsilon(10860)$ and $\Upsilon(11020)$ can be radial excitations of the $\Upsilon$, while a mass compatible with the one of $\Upsilon(10753)$ can not be found in the table.





\begin{table}[h!]
  \centering
   \footnotesize{
 \renewcommand{\arraystretch}{1.}%
 {%
 \begin{tabular}{cccccccccc}
 \cline{3-10}
  \multicolumn{1}{c}{} & \multicolumn{1}{c}{} &  \multicolumn{2}{c}{$J=0$, $Q=c$} & \multicolumn{2}{c}{$J=1$, $Q=c$}&  \multicolumn{2}{c}{$J=0$, $Q=b$} & \multicolumn{2}{c}{$J=1$, $Q=b$} \\
 \hline
  & Level  & Mod. & Exp.   & Mod.  & Exp.   & Mod. & Exp.   & Mod.  & Exp.  \\
  \hline
 $Q\bar q$ & $1S$ & 1.862 & 1.867$^*$ & 2.027 & 2.009$^*$  & 5.198 & 5.279$^*$  & 5.288 & 5.325$^*$  \\
 & $2S$  & 2.486 &  2.549   &  2.597  & 2.627 & 5.762  &  & 5.822 &  \\
 & $3S$  &    2.923  &   &  2.987 & & 6.183  & &  6.224 &  \\
 \hline
 $Q\bar s$ & $1S$  & 1.973 & 1.968$^*$   & 2.111 & 2.112$^*$ & 5.301 & 5.367$^*$  & 5.364 & 5.415$^*$ \\
 & $2S$  &  2.585 &  2.591 & 2.675 & 2.714&  5.860  &  &  5.899 &  \\
 & $3S$    & 3.018  &   & 3.069 & & 6.272   & & 6.301 &   \\
 \hline $Q\bar Q$  & $1S$   &  2.990 & 2.984$^*$  & 3.125 & 3.097$^*$& 9.387 & 9.399 & 9.405 & 9.460$^*$   \\
 & $2S$  &  3.600 & 3.637$^*$  &  3.659 & 3.686$^*$ & 10.037  & 9.999 & 10.041 & 10.023$^*$ \\
 & $3S$  &  4.006 & & 4.054 &  & 10.370 &  &   10.373 & 10.355$^*$ \\
 & $4S$  & 4.291  & & 4.308 &  & 10.620 & &    10.621 & 10.579$^*$  \\
 & $5S$  & 4.659  & & 4.735 & & 10.898 & & 10.901 &   \\
 & $6S$  & 5.549 & & 5.622 && 11.029  & & 11.029 &   \\
 \hline
$Q\bar b$ & $1S$ & 6.310 & 6.274$^*$ &  6.338 &&&&  \\
& $2S$  & 6.872  & 6.871 &   6.882 &&&& \\
& $3S$  & 7.225  &   &      7.232 &&&&  \\
 \hline
 \end{tabular}
 }
 }
 \caption{Meson masses (GeV) obtained in Ref. \cite{Carlucci:2007um}. $^*$ denotes a mass used as input to fit the parameters of the model. The first columun describes the meson quark content. The columns ``Mod'' contain results of the present model \cite{Carlucci:2007um}, the columns ``Exp.'' contain data from \cite{pdg}. }
 \label{tab:mesons}
 \end{table}


To compute diquark masses, Eq.~\eqref{eq:salpeter} is solved with the same potential introduced for mesons with a factor $1/2$ \cite{Carlucci:2007um,Richard:1983tc}. This assumption comes from the fact that, in the one-gluon-exchange approximation, the attractive potential between two quarks is half the one between a quark and an antiquark.
We only consider diquarks in the $\bar 3_c$ color representation.
The masses of diquarks with spin 0 ($[qq^\prime]$ ) and spin 1 ($\{qq^\prime\}$) are shown in Ref.~\cite{Carlucci:2007um}. 


To compute tetraquark masses we then solve Eq. \eqref{eq:salpeter} with $m_1$ and $m_2$ diquark masses. To take into account the interaction between extended objects in tetraquarks, we use the meson potential modified by a convolution with diquark wavefunctions ($\psi_d$):
\begin{equation}
  \tilde V(R)=\frac{1}{N}\int
d{\bf r_1}\int d{\bf r_2}|\psi_d({\bf r_1})|^2|\psi_d({\bf r_2})|^2
V\Big(\Big|{\bf R}+{\bf r_1}-{\bf r_2}\Big|\Big)\,.
\end{equation}

In Ref.~\cite{Carlucci:2007um}  the results for tetraquarks with hidden charm and bottom are shown. 
The masses predicted by the model are compatible with the mass of the $1^{++}$ $\chi_{c1}(3872)$ and of the $1^{+-}$ charged state $Z_c(3900)^\pm$ in the charm sector. 
It is also supported the hypothesis that $Z(4430)$ can be interpreted as the first radial excitation of the $1^{+-}$ state, which in this model has mass 4.42 GeV. 
The masses of $bq\bar b\bar q$ $1^{+-}$ states are $\sim10.3$ GeV ($1S$) and $\sim10.8$ GeV ($2S$), so the masses of $Z_c(4200)^\pm$ (observed by Belle \cite{Belle:2014nuw}), $Z_b(10610)^\pm$, $Z_b(10650)^\pm$ are not compatible with a diquark-antidiquark content in this model. 
There is however another possible excited state that can be analysed, in which one diquark is excited ($2S$) while the final tetraquark wavefunction is the $1S$ one. In the charm sector we find that these  $1^{+-}$ states have mass in the range $4.19-4.23$ GeV, and in the bottom sector $\sim10.6$ GeV.


In this paper the masses of $Qs\bar Q \bar q$ and $QQ\bar Q \bar Q$ tetraquarks are computed, and the results are shown in Table~\ref{tab:strangetet} and Table~\ref{tab:cccc}.

The masses of $Qs\bar Q \bar q$ tetraquarks are collected in Table~\ref{tab:strangetet}.
In this sector, there are two experimental exotic candidates: $Z_{cs}(4000)\, (1^+)$, with mass $3980-4010$ MeV \cite{pdg}, and $Z_{cs}(4220)^+\, (1^+)$, with mass $4216^{+50}_{-40}$ MeV \cite{LHCb:2021uow}. Only the first one has a mass compatible to the predicted values.
The masses in Table~\ref{tab:strangetet} are compatible with the masses found in \cite{ferrettisantopinto} in a compact tetraquark model.

\begin{table}[h!]
  \centering
{\footnotesize
\begin{center}
\renewcommand{\arraystretch}{1.}%
{
\begin{tabular}{cccccc}
\cline{3-6}
 & & \multicolumn{2}{c}{$Q=c$} & \multicolumn{2}{c}{$Q=b$} \\
 \hline
 $J^{P}$&Flavor content&  $1S$ & $2S$ & $1S$ & $2S$ \\
 \hline
 $0^{+}$&$[Qs][\bar Q\bar q]$& 3.961 & 4.483  & 10.283 & 10.868\\
  $1^{+}$&$[Qs]\{\bar Q\bar q\}$ & 4.008 & 4.528 & 10.321 & 10.906 \\
  $1^{+}$&$\{Qs\}[\bar Q\bar q]$ & 3.999 & 4.519 & 10.311 & 10.896 \\
 $0^{+}$&$\{Qs\}\{\bar Q\bar q\}$& 3.840 & 4.491 & 10.280 & 10.918\\
 $1^{+}$&$\{Qs\}\{\bar Q\bar q\}$& 3.952 & 4.527 & 10.312 & 10.925\\
 $2^{+}$&$\{Qs\}\{\bar Q\bar q\}$& 4.113 & 4.599 & 10.384 & 10.944\\
\hline
\end{tabular}
}
\end{center}
}
\caption{New results on $Qs\bar Q \bar q$ masses (GeV). $[Qq]$ indicates a diquark with spin 0, $\{Qq\}$ indicates a diquark with spin 1. Tetraquark spin-parity is indicated in the first column. Columns 3 and 5 contain ground-state masses  ($1S$), columns 4 and 6 contain masses of the first radial excitation ($2S$).}
 \label{tab:strangetet}
 \end{table}


Let us consider tetraquarks with only heavy quarks, the masses of which are in Table~\ref{tab:cccc}. 
The LHCb collaboration has observed structures with $cc\bar c\bar c$ quark content, among which a narrow structure $X(6900)$ with mass $6886\pm 16$ MeV \cite{LHCb:2020bwg}. 
In the present model spin splittings are very small, also in the charm sector, so the states in each column are almost degenerate, and it is not possible to assign a spin to these states from the masses. We suggest that the observed narrow structure could be a $2S$ radial excitation of the $cc\bar c \bar c$ tetraquark.
The masses in Table~\ref{tab:cccc} are a bit higher than the values found in \cite{bedolla2020}, however for some states differences are very tiny.
For comparison with results found in other models see \cite{bedolla2020}.

\begin{table}[h!]
  \centering
{\footnotesize
\begin{center}
\renewcommand{\arraystretch}{1.}%
{
\begin{tabular}{cccccccc}
\cline{2-7}
 & \multicolumn{3}{c}{$Q=c$} & \multicolumn{3}{c}{$Q=b$}\\
\hline
 $J^{PC}$ &$1S$  & $2S$   & $3S$    & $1S$  &  $2S$ & $3S$    \\
\hline 
$0^{++}$ & 6.158 & 6.761 & 7.105 & 18.798 & 19.573 & 19.893\\
$1^{+-}$ & 6.189 & 6.766 & 7.108 & 18.804 & 19.574 & 19.894\\
$2^{++}$ & 6.314 & 6.811 & 7.138 & 18.817 &  19.577 & 19.895\\
\hline
\end{tabular}
}
\end{center}
}
\caption{New results on $QQ\bar Q \bar Q$ masses (GeV) for ground state ($1S$) and radial excitations ($2S$ and $3S$). All tetraquarks are obtained from two $\{QQ\}$ diquarks, the total spin is indicated in the first column.}
 \label{tab:cccc}
 \end{table}

The model has been also applied to pentaquarks, comprising four quarks and an antiquark \cite{Giannuzzi:2019esi}. The model can only deal with two-body interactions, so pentaquarks have been studied by considering three possible ways of combining five objects two by two, as shown in  \cite{Giannuzzi:2019esi}. 
Each partial combination has to be in a $3_c$ or $\bar 3_c$ representation of the $SU(3)$ color group, while the last one has to be a singlet. 
The first combination is the diquark-diquark-antiquark model: at first two diquarks are formed, then combined to form a four-quark $3_c$ object and finally combined with the antiquark forming the colourless pentaquark. 
In the second case a diquark is at first formed, then combined, in series, with the antiquark, one quark and the last quark. The last combination is the triquark-diquark model. The first and the last combinations can only be used for pentaquarks containing at least two heavy quarks, while the second one can be used also when only one heavy quark is present or when there are only a heavy quark and a heavy antiquark.
The mass of the objects formed at each step is obtained by solving Eq. \eqref{eq:salpeter}, and the potential used has a factor $1/2$ if the interaction is between two $3_c$ or $\bar 3_c$ objects, and it is obtained from a convolution with the wavefunction of the extended object.

The masses of $QQ^\prime qq\bar{q}$ pentaquarks in model $\mathcal{A}$, $\mathcal{B}$, and $\mathcal{C}$, and of $Q qqq\bar{Q}$ in $\mathcal{B}$ are collected in \cite{Giannuzzi:2019esi}.  
Comparing the results obtained for $QQqq\bar q$ in each model, one can notice that the masses in model $\mathcal{B}$ are slightly higher than in model $\mathcal{A}$ (mean difference $\sim 0.08\%$ for $Q=c$), and those in model $\mathcal{C}$ are  slightly higher than in model $\mathcal{B}$ (mean difference $\sim 1.4\%$ for $Q=c$). In general, differences are small.

For $c qqq\bar{c}$ pentaquarks, we have found that states with spin $1/2$ have masses in the range 4.57-4.65 GeV, states with spin $3/2$ have masses in the range 4.64-4.71 GeV, the state with spin $5/2$ has mass 4.76 GeV, so they are all heavier than the resonances observed by LHCb, for which a definite spin-parity assignment is absent \cite{LHCb:2015yax,LHCb:2019kea}.
This discrepancy could be understood realising that constituent quark masses in potential models can get different values in baryon and meson spectroscopy \cite{Giannuzzi:2009gh,Maiani:2004vq}.
A rapid way of testing this hypothesis is to imagine that the constant term $V_0$ of the potential has for baryons and pentaquarks a value that is different from the one fitted by meson masses. 
Assuming the mass of the lightest spin-1/2 pentaquark with hidden charm is 4312 MeV, the value of $V_0$ reproducing such a mass is $V_0 = -0.594$ GeV, giving a mass of the heaviest spin-1/2 pentaquark equal to 4.39 GeV and a mass of the heaviest spin-3/2 pentaquark equal to 4.45 GeV, getting a better agreement with experimental observations. 
Spin splittings are not modified by a change in $V_0$, while a different value of quark masses could also affect the spin-spin interaction, so a proper investigation of this aspect would deserve a dedicated study based on baryon spectroscopy.

\section{Conclusions}
We have used the $Q\bar Q$ potential computed in AdS/QCD to study heavy hadron spectroscopy in a potential model. We have tried to adapt the two-body problem to states different from a meson, and to accomodate more complex bound states like tetraquarks and pentaquarks in the same framework, considering them as emerging from, respectively, two and three subsequent interactions.  
This model can also be used to compute the width of the analysed states  \cite{Giannuzzi:2008pv}, which, together with the mass, can give a very useful information to study hadron spectroscopy.

\acknowledgments
The work is carried out within the INFN project (Iniziativa Specifica) SPIF.


\begin{thebibliography}{99}
\bibitem{Carlucci:2007um}
\BY{M.~V.~Carlucci \textit{et al.}}
\IN{Eur. Phys. J. C}{57}{2008}{569-578}.
\bibitem{Giannuzzi:2019esi}
\BY{F.~Giannuzzi}
\IN{Phys. Rev. D}{99}{2019}{094006}.
\bibitem{pdg}
\BY{R.~L.~Workman \textit{et al.} [Particle Data Group]}
\IN{PTEP}{2022}{2022}{083C01}.
\bibitem{LHCb:2021uow}
\BY{R.~Aaij \textit{et al.} [LHCb]}
\IN{Phys. Rev. Lett.}{127}{2021}{082001}.
\bibitem{LHCb:2020bwg}
\BY{R.~Aaij \textit{et al.} [LHCb]}
\IN{Sci. Bull.}{65}{2020}{1983-1993}.
\bibitem{Andreev:2006ct}
\BY{O.~Andreev and V.~I.~Zakharov}
\IN{Phys. Rev. D}{74}{2006}{025023}.
\bibitem{Giannuzzi:2009gh}
\BY{F.~Giannuzzi}
\IN{Phys. Rev. D}{79}{2009}{094002}.
\bibitem{Maiani:2004vq}
\BY{L.~Maiani, F.~Piccinini, A.~D.~Polosa and V.~Riquer}
\IN{Phys. Rev. D}{71}{2005}{014028}.
\bibitem{other}
\BY{R.~L.~Jaffe}
\IN{Phys. Rept.}{409}{2005}{1-45};
\BY{R.~F.~Lebed}
\IN{Phys. Lett. B}{749}{2015}{454-457};
\BY{L.~Maiani, A.~D.~Polosa and V.~Riquer}
\IN{Phys. Lett. B}{749}{2015}{289-291};
\BY{M.~N.~Anwar \textit{et al.}}
\IN{Eur. Phys. J. C}{78}{2018}{no.8, 647}.
\bibitem{ferrettisantopinto}
\BY{J.~Ferretti and E.~Santopinto}
\IN{JHEP}{04}{2020}{119}.
\bibitem{bedolla2020}
\BY{M.~A.~Bedolla, J.~Ferretti, C.~D.~Roberts and E.~Santopinto}
\IN{Eur. Phys. J. C}{80}{2020}{no.11, 1004}.
\bibitem{Barnes:2005pb}
\BY{T.~Barnes, S.~Godfrey and E.~S.~Swanson}
\IN{Phys. Rev. D}{72}{2005}{054026}.
 \bibitem{Colangelo:1990rv}
\BY{P.~Colangelo, G.~Nardulli and M.~Pietroni}
\IN{Phys. Rev. D}{43}{1991}{3002}.
\bibitem{Richard:1983tc}
\BY{J.~M.~Richard and P.~Taxil}
\IN{Phys. Lett. B}{128}{1983}{453-456}.
\bibitem{Belle:2014nuw}
\BY{K.~Chilikin \textit{et al.} [Belle]}
\IN{Phys. Rev. D}{90}{2014}{112009}.
\bibitem{LHCb:2015yax}
\BY{R.~Aaij \textit{et al.} [LHCb]}
\IN{Phys. Rev. Lett.}{115}{2015}{072001}.
\bibitem{LHCb:2019kea}
\BY{R.~Aaij \textit{et al.} [LHCb]}
\IN{Phys. Rev. Lett.}{122}{2019}{222001}.
\bibitem{Giannuzzi:2008pv}
\BY{F.~Giannuzzi}
\IN{Phys. Rev. D}{78}{2008}{117501}.
  

  
  \end{thebibliography}
\end{document}